\documentclass{aa}  

\usepackage{graphicx}
\usepackage{txfonts}
\usepackage{rotating}
\usepackage{answers}
\usepackage{xcolor}
\usepackage{scalefnt}
\usepackage{xcolor}
\usepackage{ragged2e}
\usepackage{float}

\usepackage[colorlinks=true, allcolors=blue]{hyperref}
\newbox\grsign \setbox\grsign=\hbox{$>$} \newdimen\grdimen \grdimen=\ht\grsign
\newbox\simlessbox \newbox\simgreatbox
\setbox\simgreatbox=\hbox{\raise.5ex\hbox{$>$}\llap
     {\lower.5ex\hbox{$\sim$}}}\ht1=\grdimen\dp1=0pt
\setbox\simlessbox=\hbox{\raise.5ex\hbox{$<$}\llap
     {\lower.5ex\hbox{$\sim$}}}\ht2=\grdimen\dp2=0pt

\newbox\simppropto
\setbox\simppropto=\hbox{\raise.5ex\hbox{$\sim$}\llap
     {\lower.5ex\hbox{$\propto$}}}\ht2=\grdimen\dp2=0pt

\begin{document}

\title{Revisited parameters for the twin bulge globular clusters NGC~6528 and NGC 6553} 
\subtitle{}
   
   \author{S. Ortolani\inst{1,2,3}
        \and 
   S. O. Souza\inst{4}
          \and
        D. Nardiello\inst{1,3}
        \and
        B. Barbuy\inst{5}
        \and
        E. Bica\inst{6}
%        \and
%        M. Griggio\inst{3,11}
%        \and
%        B. Dias\inst{12}
                  }
\institute{Universit\`a di Padova, Dipartimento di  Fisica e Astronomia, Vicolo dell'Osservatorio 2, I-35122 Padova, Italy
\and
Centro di Ateneo di Studi e Attivit\`a Spaziali “Giuseppe Colombo” – CISAS, Via Venezia 15, 35131 Padova, Italy
\and
INAF-Osservatorio di Padova, Vicolo dell'Osservatorio 5, I-35122 Padova, Italy
\and
Max Planck Institute for Astronomy, K\"onigstuhl 17, D-69117 Heidelberg, Germany 
\and
Universidade de S\~ao Paulo, IAG, Rua o Mat\~ao 1226, Cidade Universit\'aria, S\~ao Paulo 05508-900, Brazil
\and
Universidade Federal do Rio Grande do Sul, Departamento de Astronomia, CP 15051, Porto Alegre 91501-970, Brazil
%\and
%Dipartimento di Fisica, Universit\`a di Ferrara, Via Giuseppe Saragat 1, %Ferrara I-44122, Italy
%\and
%Instituto de Astronom\'ia, Universidad Nacional Aut\'onoma de M\'exico, %A. P. 106, C.P. 22800, Ensenada, B. C., M\'exico
}

   \date{Received ; accepted}

% \abstract{}{}{}{}{}
% 5 {} token are mandatory
 
  \abstract
   {NGC~6528 and NGC~6553 are among the most metal-rich globular clusters in the Galactic bulge. They represent the upper end of the chemical enrichment in the Galaxy, and can inform on the processes of
   cluster formation and enrichment.}
   {We aim to refine the fundamental parameters of NGC~6528 and NGC~6553, based on proper motion-corrected 
   \emph{Hubble Space Telescope} WFC3  and ACS photometries.}
   { In order to derive the fundamental parameters age, distance, reddening, and the total-to-selective  absorption coefficient,  we employed a Bayesian isochrone fitting.  Age and metallicity are mainly constrained by the turn-off morphology,
   thanks to the unprecedented quality of the proper-motion cleaned photometry.}
   {The two clusters show  remarkably similar Colour-Magnitude Diagrams.
   We derived an age of 11$\pm$0.5 Gyr  with a solar metallicity for both clusters. 
   The reddening for NGC~6528 and NGC~6553 is E(B-V) = 0.63 and 0.76 and the
   distances from the Sun are d$_{\odot}$ = 7.85 and 5.1 kpc, respectively,  recalling that distances strictly
   depend on the adopted total-to-selective absorption parameter.
   }
   {The age of these metal-rich clusters is about 2 Gyr younger than the moderately metal-poor bulge clusters.
   The ages and metallicities are remarkably identical to that of the bulk of bulge field stars.}

    
   \titlerunning{The twin globular clusters NGC~6528 and NGC~6553}
   \authorrunning{Ortolani et al.}
   \maketitle
%
%-------------------------------------------------------------------
    
\section{Introduction}

The high-metallicity of the globular clusters (GCs) NGC~6528 and NGC~6553 was
identified by \citet{vandenbergh79} and \citet{hartwick75} respectively, from
early Colour-Magnitude Diagrams (CMD). This was further confirmed in 
integrated light by  \citet{zinn80}, \citet{bica83}, \citet{zinnwest84}, 
\citet{armandroff88}. 
Their high metallicities became more clear with the library of integrated spectra by
\citet{bica86}. To accomplish these observations, in order to avoid bright cool M giants that could
mask the spectra, the telescope was moved perpendicularly to the slit, and the reduced spectra
were then added in alpha (RA). From this survey the most metal-rich clusters were found to be
 NGC~6528, NGC~6553, and NGC~6440.
\citet{bica88} showed that the spectra of 
metal-rich clusters were very similar to spectra of ellipticals and bulges of spirals.

Other bulge globular clusters appear to have even higher metallicity, such as Terzan~5 and Liller~1.
\citet{zinnwest84} found [Fe/H]=+0.24 for Terzan 5, confirmed by \citet{ortolani96} to show
a solar-like metallicity.
Later, \citet{frogel95} suggested that Liller~1 would be the most metal-rich globular cluster in the
Galaxy. Recently, \citet{alvarez-garay24} deduced metallicities of [Fe/H] = -0.22 and +0.22 for the two
main populations of this cluster. We opted to further study NGC 6528 and NGC~6553 because they have
a much lower reddening and lower field contamination than Terzan~5 and Liller~1.
In addition, several {\it Hubble Space Telescope} \emph{(HST)} high resolution images are available in the Space Telescope Science Institute (STScI) archive covering the last 20 years which allows us obtaining proper motion estimation, and therefore, a cleaned cluster CMD.

Following these identifications,  a plethora of studies on NGC~6528 and NGC~6553
as the central object of the work,
were presented in the literature.
Among the most specific ones, we can cite
\citet{ortolani92}, \citet{sagar95},  \citet{ortolani95}, \citet{richtler98}, 
\citet{davidge00}, \citet{coelho01}, \citet{feltzing02}, 
\citet{momany03}, \citet{zoccali04}, \citet{origlia05},
\citet{lagioia14}, 
\citet{calamida14}, \citet{dias15}, \citet{liu17} and
\citet{munoz21} for NGC~6528; and
\citet{ortolani90}, \citet{barbuy92}, \citet{demarque92},
\citet{sagar95},
\citet{guarnieri97}, \citet{guarnieri98}, 
\citet{barbuy99b}, \citet{sagar99},  \citet{cohen99}, \citet{coelho01},
\citet{vallenari01},  \citet{beaulieu01}, \citet{zoccali01},
\citet{coelho01}, \citet{origlia02}, \citet{melendez03}, 
\citet{alvesbrito06}, \citet{dias15}, \citet{tang17},
\citet{munoz20}, and \citet{montecinos21} for NGC~6553.

Additionally  \citet{guillot11} identified an X-ray binary, and
\citet{minniti15} a black hole candidate in the field of NGC~6553.

These two clusters are included in the list of in situ bulge globular clusters by
 \citet{bica16,bica24} and this classification is confirmed from  orbital analyses \citep{massari19,perez-villegas20}, and likewise by  \cite{belokurov24}.
 \citet{callingham22} instead assigns NGC~6553 to Kraken \citep{kruijssen20}, and NGC~6528 to the
 main bulge.
 
In the present work, we carry out a Colour-Magnitude Diagram  
study of the globular clusters NGC~6528 and NGC~6553. 

%\multicolumn{2}{c}{\hbox{ NGC~6528}  & \multicolumn{2}{c}{\hbox{ NGC~6553}  \\  
The clusters are projected towards the Galactic bulge, with equatorial coordinates
$\alpha = 18^{\rm h}10^{\rm m}18.4^{\rm s}$, $\delta = -30^{\rm o}03'20.8''$ 
and Galactic coordinates l, b = 1.14$^{\circ}$, -4.17$^{\circ}$ for NGC~6528, and
$\alpha = 18^{\rm h}09^{\rm m}15.68^{\rm s}$, $\delta = -25^{\rm o}54'27.9''$,
and l, b = 
 5.25$^{\circ}$, -3.02$^{\circ}$ \citep[2010 edition]{harris96} for NGC~6553. 
 NGC~6528 is in Baade's Window.

In Section \ref{sec:2} the observations and data reduction are described.  In Section \ref{sec:3}
parameters  from the literature are reviewed, and
 isochrone fitting is applied to CMDs and described. 
 Chemical abundances are discussed in \ref{sec:4}. Concluding remarks are given in Section \ref{sec:disc}.

%--------------------------------------------------------------------

%--------------------------------------------------------------------

\section{Observations and data reduction}
\label{sec:2}

In this work, we used HST observations of the two globular clusters NGC\,6528 and NGC\,6553. For NGC\,6528 we analyzed images in ACS/WFC F606W and F814W filters collected during GO-9453 (PI: Brown, epoch: 2002.60)\citep{brown02}, and in WFC3/UVIS F390W, F555W, and F814W filters obtained during GO-11664 (PI: Brown, epoch: 2010.55)\citep{brown10}. For NGC\,6553, we used ACS/WFC images collected during GO-10573 (PI: Mateo, epoch: 2006.26)\citep{mateo05} in F435W, F555W, and F814W filters, WFC3/UVIS data in F555W and F814W filters during GO-15232 (PI: Ferraro, epoch: 2017.81)\citep{ferraro17}, and WFC3/IR images in F110W and F160W obtained during GO-16282 (PI: Correnti, epoch: 2020.81)\citep{correnti20}. A log of the observations is reported in Table~A\ref{obs}.

For the data reduction and the extraction of astro-photometric catalogues of the two clusters, we followed the procedure adopted by \citet{nardiello18,nardiello18b}.
First, we extracted astro-photometric catalogs of each image by using the software \texttt{hst1pass} (first-pass photometry, \citealt{2022wfc..rept....5A}), that makes use of perturbated library PSFs to fit the positions and magnitude of the stars in the images. For each filter, six-parameters transformations were used to match the positions of the stars measured in each image, while linear photometric transformations were used to transform the magnitudes of the stars measured in each image to a common reference system (we adopted the longest exposure image as reference). We combine all the transformations to obtain a first-guess single catalogue that contains averaged positions and fluxes calculated in a defined reference system. We used perturbated PSFs, transformations and images to carry on the second-pass photometry with KS2 (\citealt{2008AJ....135.2114A}): this software performs a  simultaneous analysis of all the images to extract precise positions and fluxes of all the stars detectable in the field through different iterations. In each iteration it subtracts the stars already measured and searches for fainter stars.  The output of the KS2 are a catalog with mean positions, fluxes, error on the fluxes, quality-of-fit parameter, sharp parameter. These parameters were used to reject bad measurements as done in \citet{scalco21}. Moreover, single catalogs associated to each image in which positions and fluxes of the stars measured in each single image are also available. We used them to compute the proper motions between the two epochs and following the method of \citet{libralato21}. This method is based on the use of local network of bona-fide cluster members to compute the mean displacements of the stars between two epochs. For each star in the catalog, the software searches for bona-fide members in the neighbourhood (usually between 20 and 50 stars) and calculate the mean displacement between the first and second epoch. Because as reference we adopted the stars in the cluster, the cluster members are centered at (0,0) in the vector-point diagram.

We corrected the CMDs for differential reddening following the procedure adopted by \citet{milone12}. Briefly, we selected those with highest probability to be cluster member on the basis of their position on the CMD and their proper motions. For each star in the catalog, we  selected from this sample the closest 50 cluster stars and measured the mean difference between their colour and the fiducial cluster sequence along the reddening direction. We assumed this mean value as an estimate of the local differential reddening.

\begin{figure*}
\centering
\includegraphics[width=8cm]{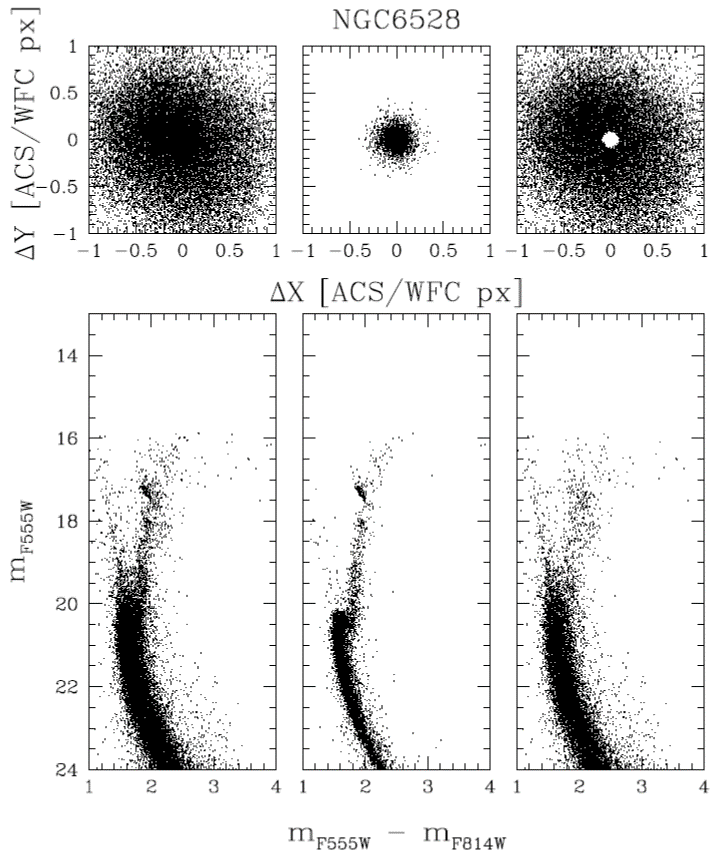}
\includegraphics[width=8cm]{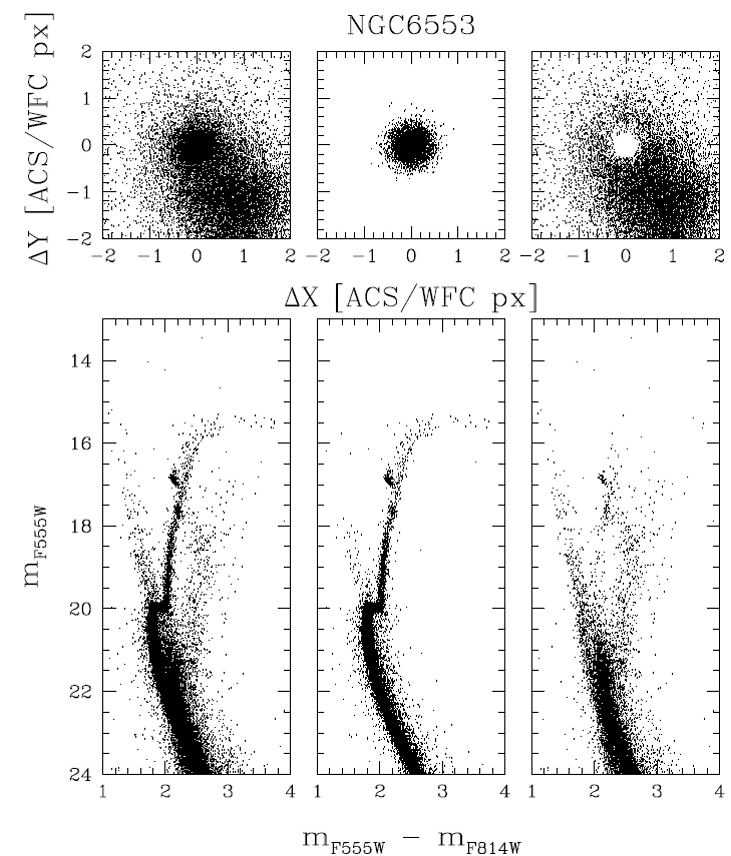}
\caption{NGC~6528 and NGC~6553: original and proper-motion cleaned CMDs  in F814W vs. F555W=-F814W.
For both clusters, left panel is the original photometry, middle panel is the proper-motion cleaned,
and the right panel is the field.} 
\label{fig1}
\end{figure*}

\begin{figure}
\centering
\includegraphics[width=\columnwidth]{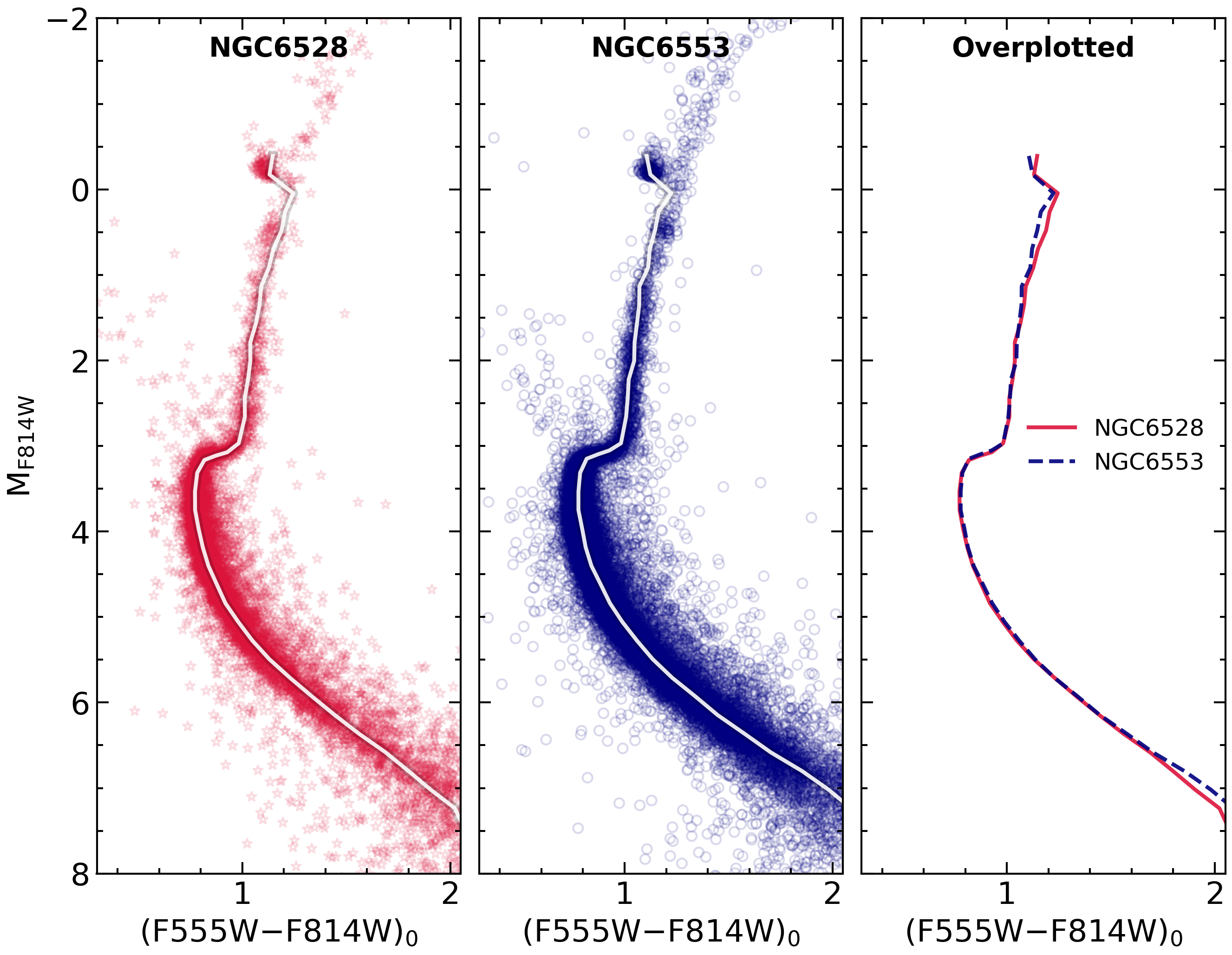}
\caption{Comparison of the CMDs of the GCs NGC 6528 and NGC 6553. The left and middle panels show the CMDs of NGC 6528 (red) and NGC 6553 (blue), respectively, with their corresponding fiducial lines overplotted in white. The right panel directly compares the fiducial lines of both clusters, with NGC 6528 shown as a red solid line and NGC 6553 as a blue dashed line. The similarity between the fiducial lines highlights the close resemblance of their stellar populations.} 
\label{fidulines}
\end{figure}

\begin{figure*}
\centering
\includegraphics[width=15cm]{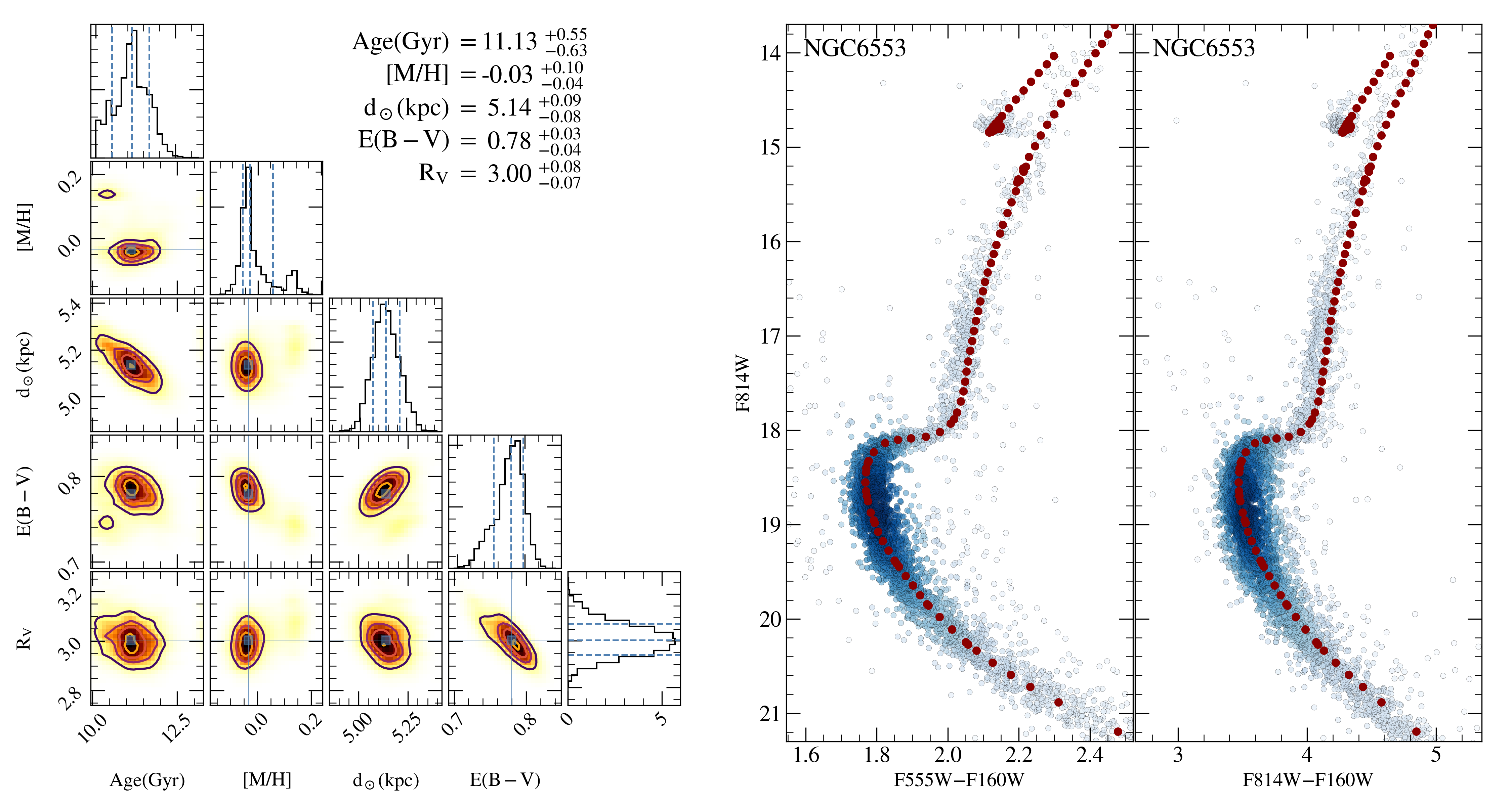}
\caption{ Isochrone fitting result for NGC6553. Left: Corner plot displaying the posterior probability distributions of the fundamental parameters obtained from isochrone fitting. The 1D marginalized distributions along the diagonal show the median values (solid lines) and the 16th–84th percentile ranges (dashed lines), while the 2D contour plots depict the covariances between parameters. The best-fit values with uncertainties are listed in the upper right corner. Right: Observed CMDs in the HST filters: F814W vs. F390W–F814W (left panel) and F814W vs. F555W–F814W (right panel). The CMD is coloured by density. The best-fit isochrone, derived from the MCMC analysis, is overlaid in red. The isochrone accurately reproduces the main sequence, subgiant branch, and red giant branch, validating the derived cluster parameters.} 
\label{n6553fit}
\end{figure*}

\begin{figure*}
\centering
\includegraphics[width=15cm]{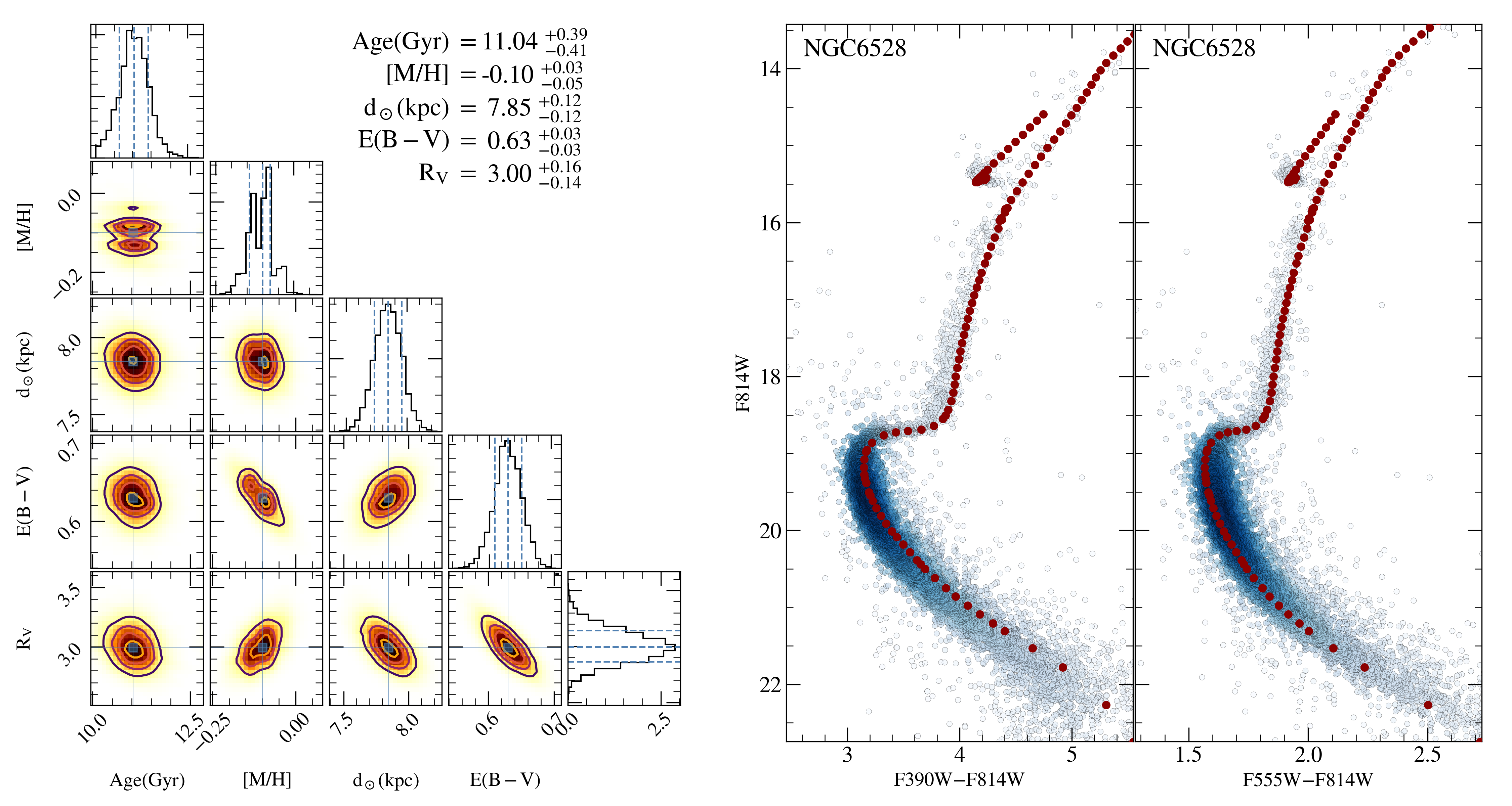}
\caption{Same as Figure \ref{n6553fit} but for NGC~6528 and CMDs in  F814W vs. F390W-F814W (middle) and F814W vs. F555W-F814W (right).} 
\label{n6528fit}
\end{figure*}

\section{Colour-Magnitude Diagrams and Isochrone fitting}
\label{sec:3}

In Table \ref{tab:literature} we report literature age, reddening E(B-V), distance, and metallicity for the two
sample clusters. A more recent review on the distances is presented by \citet{baumgardt21},
where distances of 7.8 and 5.3  kpc respectively for the clusters are given, by averaging the results
from optical, near-infrared and Gaia  EDR3 \citep{gaia21}.

\begin{table}
\small
\scalefont{0.9}
\caption[4]{\label{tab:literature}
Literature reddening, ages, distance and metallicities for NGC~6528 and NGC~6553.}
\begin{tabular}{l@{}r@{}r@{}r@{}c@{}c@{}}
\hline
\noalign{\smallskip}
\hbox{E(B-V)} & \hbox{\phantom{-}Age} & \hbox{\phantom{-}d$_{\odot}$} & \hbox{\phantom{-}[Fe/H]}   & \hbox{Method} &\hbox{Reference}  \\
\hbox{} &\hbox{\phantom{-}Gyr} & \hbox{kpc} & \hbox{} & \hbox{} & \hbox{}   \\ 
\noalign{\smallskip}
\hline
\noalign{\smallskip}
\noalign{\hrule\vskip 0.1cm}
%\multicolumn{4{c}{\hbox{ NGC~6528}   \\  
     &    &    & \bf NGC~6528 & \\
\noalign{\smallskip}
\noalign{\hrule\vskip 0.1cm}
\noalign{\smallskip}
0.52$\pm$0.07 & --- & 6.1& \phantom{-}$-$0.4 & CMD & VDB79 \\  
0.56$\pm$0.03 & --- & --- & \phantom{-}+0.01$\pm$0.10& IPhot & Z80 \\  
--- & --- & --- & \phantom{-}$-$0.18$\pm$0.36 & IPhot & B83 \\  
--- & --- & --- & \phantom{-}+0.11$\pm$0.21 & spec & ZW84 \\  
--- & --- & --- & \phantom{-}$-$0.23$\pm$0.11 & IS & A88 \\ 
0.66$\pm$0.09 & --- & --- & \phantom{-}$-$0.05$\pm$0.2  & IS & B86\\  
0.55$\pm$0.1 & 14 & 7.5 & \phantom{-}high &  CMD & O92 \\  
---  & 11$\pm$1.5 & --- & \phantom{-}+0.00  & CMD & O95\\     
0.52$\pm$0.05 & --- & 7.83 & \phantom{-}$-$0.1$\pm$0.1 & CMD & B98 \\ 
0.8 & .---& ---  & \phantom{-}+0.1 & CMD & R98 \\
0.29 & .---& 0.7$\pm$1.1 & \phantom{-}+0.0$\pm$0.09   & CMD & H99 \\ 
-- & --- & --- & \phantom{-}+0.07$\pm$0.02 & spec & C01 \\
--- & 11$\pm$0.2 & 7.2 & \phantom{-}+0.00 & CMD & F02 \\
0.55 & 12.6 & 7.7 & \phantom{-}+0.00 & CMD & Mo03 \\
0.46 & --- & --- & \phantom{-}$-$0.1$\pm$0.2 & spec & Z04\\
0.62 & --- & --- & \phantom{-}$-$0.17$\pm$0.01 & spec & O05 \\
0.54 & --- & 7.9 & \phantom{-}$-$0.1 3$\pm$0.05 & spec &  D15\\
0.54 & 11$\pm$1 & 7.9 &  \phantom{-}+0.20 & CMD & L14 \\
-- & --- & --- & \phantom{-}+0.04$\pm$0.07 & spec & L17 \\
--- & --- & --- & \phantom{-}$-$0.19$\pm$0.04 & spec & S17 \\
--- & --- & --- & \phantom{-}$-$0.20$\pm$0.06 & spec & M18 \\
-- & --- & 7.83$\pm$0.23 &  --  & Gaia & B21 \\
\noalign{\smallskip}
\noalign{\hrule\vskip 0.1cm}
\noalign{\smallskip}
     &    &    & \bf NGC~6553 & \\
     \noalign{\smallskip}
\noalign{\hrule\vskip 0.1cm}
\noalign{\smallskip}
---           & --- & 4.8 & \phantom{-}high           & CMD &  H75 \\ 
0.78$\pm$0.03 & --- & --- & \phantom{-}+0.04$\pm$0.10 & IPhot & Z80 \\ 
--- & --- & --- & \phantom{-}+0.47$\pm$0.63  & IPhot &  B83\\    
--- & --- & --- & \phantom{-}$-$0.29$\pm$0.11 & spec & ZW84\\  
0.79$\pm$0.09 & --- & --- &  \phantom{-}+0.10$\pm$0.4 & IS & B86\\
1.08 & 11.5$\pm$2.5&  --- & \phantom{-}high &  CMD & O90\\ 
0.8 & ---& ---  & \phantom{-}$-$0.2$\pm$0.2 & spec & B92 \\
--- & 11$\pm$1.5 & ---  & \phantom{-}+0.00  & CMD & O95\\ 
--- & 8-13 & --- &  \phantom{-}+0.00 & CMD & D92\\
0.7$\pm$0.05 & --- & 5.25 &  \phantom{-}$-$0.22$\pm$0.05  & CMD/spec & G97,G98 \\
0.7$\pm$0.06 & --- & 5.1 & \phantom{-}$-$0.2$\pm$0.1 & CMD & B98 \\
0.48 & ---&  6.2$\pm$0.7 & \phantom{-}$-$0.04$\pm$0.08 & CMD & H99 \\ 
0.7& --- & 5.1 & \phantom{-}$-$0.55$\pm$0.2 & spec & B99 \\
0.59$\pm$0.0 &  --- & 5.0 & \phantom{-}$-$0.1$\pm$0.1 &  CMD  & S99 \\ 
--- & --- & --- & \phantom{-}$-$0.16$\pm$0.2 & spec & C99\\
0.63$\pm$0.03 & 13$\pm$1 & 5.3 & \phantom{-}$-$0.10 & CMD & Z01\\ 
0.72 &  13  & ---   & \phantom{-}$-$0.30  &  CMD  & B01 \\
--- & --- & --- & \phantom{-}$-$0.16$\pm$0.08 & spec & C99\\
--- & --- & --- & \phantom{-}$-$0.3$\pm$0.2 & spec & O02\\
0.7 & --- & 6.0 & \phantom{-}$-$0.2$\pm$0.15 & spec & A06 \\
0.63 & --- & --- & \phantom{-}$-$0.13$\pm$0.02& spec & D15 \\
--- & --- & --- & \phantom{-}$-$0.15$\pm$0.05 & spec & T17 \\
--- & --- & --- & \phantom{-}$-$0.11$\pm$0.04 & spec & S17 \\
--- & --- & --- & \phantom{-}$-$0.10$\pm$0.01 & spec & M21\\
-- & --- & 5.33$\pm$0.13  &  --  & Gaia & B21 \\
\noalign{\hrule\vskip 0.1cm}               
\end{tabular}
\tablebib{
\citet[][A06]{alvesbrito06}; 
\citet[][A88]{armandroff88};  \citet[][B83]{bica83}; \citet[][B86]{bica86}; 
\citet[][B92]{barbuy92}; \citet[][B99]{barbuy99b};  \citet[][B21]{baumgardt21};
\citet[][B01]{beaulieu01};  \citet[][C99]{cohen99}; \citet[][C01]{carretta01};
\citet[][D92]{demarque92};
\citet[][D15]{dias15};  \citet[][F02]{feltzing02}; \citet[][G97]{guarnieri97}; \citet[][G98]{guarnieri98};
\citet[][H75]{hartwick75}; \citet[][H99]{heitsch99};  
\citet[][L14]{lagioia14}; \citet[][L17]{liu17}; \citet[][Me03]{melendez03};
\citet[][Mo03]{momany03}; \citet[][M18]{munoz18}; \citet[][M21]{montecinos21};
\citet[][O02]{origlia02}; \citet[][O05]{origlia05};
\citet[][O90]{ortolani90};  \citet[][O92]{ortolani92};  \citet[][O95]{ortolani95};
 \citet[][R98]{richtler98} 
\citet[][S99]{sagar99}; \citet[][S17]{schiavon17}; \citet[][T17] {tang17};
\citet[][VDB79]{vandenbergh79};
\citet[][Z80]{zinn80};  \citet[][ZW84]{zinnwest84};  \citet[][Z04]{zoccali04}.
Methods for age, distance, and metallicity derivation: IPhot: integrated photometry; 
IS: integrated spectra;
CMD: isochrone fitting on CMDs; \emph{Gaia} proper motions, spec:spectroscopy.}
\end{table}

The fundamental parameters — age, metallicity, distance, reddening, and total-to-selective extinction ratio, R$_V$,  were derived using the SIRIUS code \citep{souza20,souza24}. The code employs the Bayesian method of Markov-chain Monte Carlo (McMC) to obtain probability distributions for each parameter by comparing the observed CMD with synthetic CMDs built from each set of parameters randomly drawn during the fitting process. The McMC was applied using the \texttt{Python} library, \texttt{emcee} \citep{emcee}, and \texttt{PyDE}\footnote{\url{https://github.com/hpparvi/PyDE}}, a global optimisation that uses differential evolution.

In order to build the synthetic diagrams, the SIRIUS code uses the PAdova and TRieste Stellar Evolution Code (PARSEC)\footnote{Version 3.8, \url{https://stev.oapd.inaf.it/cgi-bin/cmd_3.8}} \citet{bressan12}. The parameter space spans ages between 7 Gyr and 14 Gyr with intervals of 0.1 Gyr and metallicities between $-2.0$ and $+0.3$ with intervals of $0.05$ dex. The code then interpolates the isochrone grid to obtain the model for each McMC realization. We assume the initial mass function from \cite{kroupa01}. All the details on how the code constructs the synthetic population can be found in \cite{souza24}. Briefly, the synthetic CMDs are constructed by drawing stellar masses from the IMF and interpolating the corresponding magnitudes from the PARSEC isochrone grid according to the parameters selected at each McMC step. A fraction of these stars are treated as unresolved binaries, with secondary components assigned using a randomly selected mass ratio, and their combined magnitudes computed by summing their fluxes. To simulate observational uncertainties, a magnitude-dependent error function derived from the observed photometric errors is applied, anchored to the position of the turn-off to preserve alignment with the observed CMD. The extinction correction is computed at each iteration using the adopted R$_V$ value and extinction coefficients interpolated from the extinction law of \citet{cardelli89}. Finally, a luminosity function is applied to the synthetic sample to reproduce the observed distribution of stars in magnitude space.

The R$_V$ value is determined by fitting simultaneously at least two different CMD for the same cluster. Usually, one CMD is composed of NIR filters to reduce the effect of reddening. This approach was first applied by \cite{pallanca21} and \cite{souza21, souza23}, and implemented as simultaneous isochrone fitting by \cite{souza24}. To improve the age determination, we give more weight to the turn-off (TO) region since the high-quality data gives us a narrow and distinct shape of the TO, being the most important feature to be fitted.

Although the SIRIUS code provides a robust statistical framework for deriving cluster parameters, several sources of systematic uncertainty remain. The adoption of a specific stellar evolutionary model incorporates assumptions about stellar physics, such as convective overshooting, mass loss, and the treatment of advanced evolutionary stages. These assumptions can lead to model-dependent biases in the recovered parameters, particularly age and metallicity \citep[see an example for the GC HP1 in][]{kerber19}. The adoption of a fixed initial mass function from \citet{kroupa01} also contributes to systematic uncertainty, especially in systems that may have experienced dynamical evolution or mass segregation. Nevertheless, the assumption of an IMF does not have an impact on our parameter derivation, given that the fitting is performed with more weight in the MSTO. An additional systematic on the extinction and R$_V$ estimates could be introduced by any residual zero-point offsets, colour terms, or spatial variations in reddening. Therefore, we stress that the parameter uncertainties presented in this work and provided by the SIRIUS code must be assumed as internal errors of the Bayesian method.

The two proper motion-cleaned  CMDs are shown in  Figure \ref{fig1} 
for the clusters, with  the F555W and F814W filters of the WFC3 camera. 
It is important to stress that the proper-motion detection allowed to be very efficient in the cleaning
from field stars.
In particular NGC~6553 has a significant proper motion leading to most of the cluster stars cleaned from the bulk of the field population. 
In Figure \ref{fidulines} the mean loci corresponding to  the CMDs shown in Figure \ref{fig1}  are compared.
Note the striking similarity between them, despite the different number of stars.
The resulting, cleaned CMDs
supersede those of the previous literature thanks to the wide time base between the first and the second epoch
(about 14 years).

In Figure \ref{n6553fit}, are presented the F814W vs. F555W-F160W and F814W vs. F814W-F160W CMDs for NGC~6553.
The choice of optical combined with infrared filters was adopted to have a wide wavelength range
and two combinations of colours allowing to derive the reddening law parameter R$_{\rm V}$.
The resulting fit yields an age of 11.1$\pm$0.5 Gyr, a reddening of E(B-V)=0.78$\pm$0.04, and a distance
from the Sun of d$_{\odot} = 5.14\pm0.09$ kpc, with R$_{\rm V}=3.0\pm0.1$.
The derived distance is in good agreement with \citet{baumgardt21}, just at the lower edge of 
their average error bar, and also compatible with the Gaia parallaxes error bar. 
This is also in agreement with most of the literature (Table \ref{tab:literature}).

 The total-to-selective absorption parameter R$_{\rm V}$ = 2.9, is somewhat lower than the standard value
 of 3.1, compatible with the location of the cluster at about $4^\circ$ below the Galactic plane.
 The metallicity of [M/H] $= -0.07$ is compatible with most of the literature giving 
 [Fe/H] $\approx -0.2$ and alpha-enhancement around [$\alpha$/Fe] $= +0.2$. The age of 11.1 Gyr is lower than the literature ages, and this is explained by the fact that previous
 estimations were based on comparisons with the halo globular cluster NGC104 (47 Tuc). It appears now
 that 47 Tuc has an age, with present data, of 11.5 Gyr \citep{gerasimov24}.  

 In Figure \ref{n6528fit} are given the F814W vs. F390W-F814W and F814W vs. F555W-F814W CMD for NGC~6528.
It is clear that the metallicity [M/H]$ = -0.10$ and age of $11.04$ Gyr of this cluster are very similar to those of NGC~6553, as expected.
The reddening E(B-V)=0.63 is similar to other literature values.
 The distance from the Sun depends on the assumed R$_{\rm V}$ value. The study of \citet{saha19}, based on
 RR Lyrae stars, indicate that the reddening parameter R$_{\rm V}$ could be lower in Baade's Window, in particular including
 NGC~6528, relative to the study by \citet{fitzpatrick99}. Since we do not have a wavelength range large enough
 to constrain the reddening law, we adopt a conservative value of R$_{\rm V} = 3.0$ as a prior in the isochrone fitting.
 The resulting distance from the Sun is d$_{\odot} = 7.85\pm0.12$ kpc.

 Therefore, these metal-rich clusters appear to be different from the moderately metal-poor clusters
 NGC 6522, NGC 6558, HP~1, AL~3, among others, that are as old as 13 Gyr - see Table 3 from \citet{souza24}.
 
The age and metallicity obtained from NGC~6553 and NGC~6528 are remarkably similar to those found from the sample of field stars in the   Sagittarius Window field in the Galactic bulge by \citet{clarkson08} from a proper motion \emph{HST} cleaned sample.  
They found that the bulk of the field population 
shows a best age from isochrone turn-off  fitting of 11 Gyr with a solar metallicity. The similarity of the two clusters with the Baade's Window field was also 
presented by \citet{ortolani95}. These results reinforce the idea that these clusters are genuine representatives of 
the bulk of the bulge field stellar population, and this result give a tighter constraint on the age and age-metallicity relation of the Galactic bulge population.

\section{Chemical abundances}
\label{sec:4}

\begin{figure*}
\centering
\includegraphics[width=15cm]{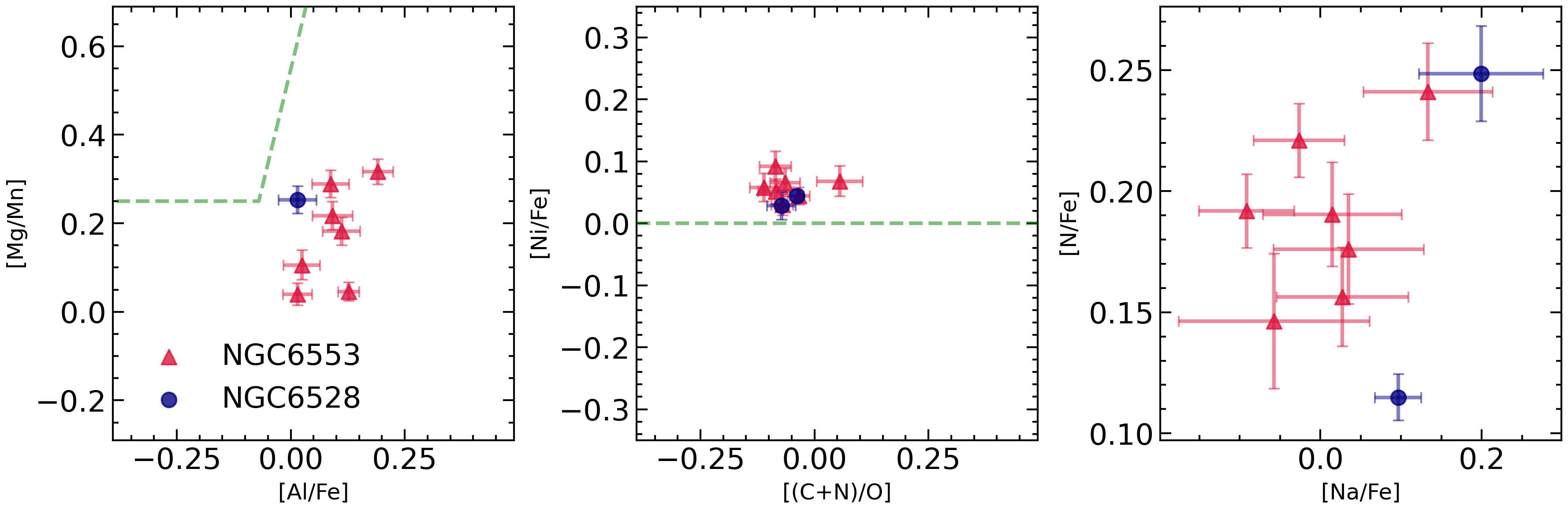}
\caption{ [Mg/Mn] vs. [Al.Fe] (left panel) , [Ni/Fe] vs. [(C+N)/O] (middle panel),
indicating in situ origin for the clusters, 
and [N/Fe] vs. [Na/Fe] (right panel), indicating that the
stars  selected are low-N and low-Na.} 
\label{insitu}
\end{figure*}

In Table \ref{abund} are reported  mean value element abundances from the literature.
It must be pointed out
that in \citet{schiavon17,schiavon24} the high abundances of N and Na in part of the stars indicate that they
correspond to a second generation (2G), and the ones that are N, and Na-normal should be first generation (1G)
stars. For this reason, we only took into account the stars with low N abundances, which should be
first generation stars.
A particularly important result is the Al abundance, which is relatively high in all the stars, pointing out to
an in-situ origin of this cluster, therefore not identified to the Kraken structure 
\citet{kruijssen20},
and contradicting \citet{callingham22}. In fact, all ex-situ objects
show low Al.
In Figure \ref{insitu} are shown indicators of in situ or ex situ origin of the clusters:
[Mg/Mn] vs. [Al/Fe] (left panel) indicates that the two clusters are found in the locus of in situ objects \citep{helmi18,limberg22,souza23};
[Ni/Fe] vs. [(C+N)/O] show Ni-relative-to-iron abundances higher than solar, again indicating an in situ origin 
according to \citet{nissen10} \citep[see also][]{barbuy24}; and [N/Fe] vs. [Na/Fe] indicating that the selected stars are N- and Na-low, therefore 1G stars.

The multiple stellar populations in the two sample clusters were made evident from photometric data
from the Blanco DECam survey of the Galactic bulge by \citet{kader22}, with approximately half of the
stars as first generation and half as second. As noted above, this is in agreement with the
relevant element abundances. 

\begin{figure}
    \centering
    \includegraphics[width=\linewidth]{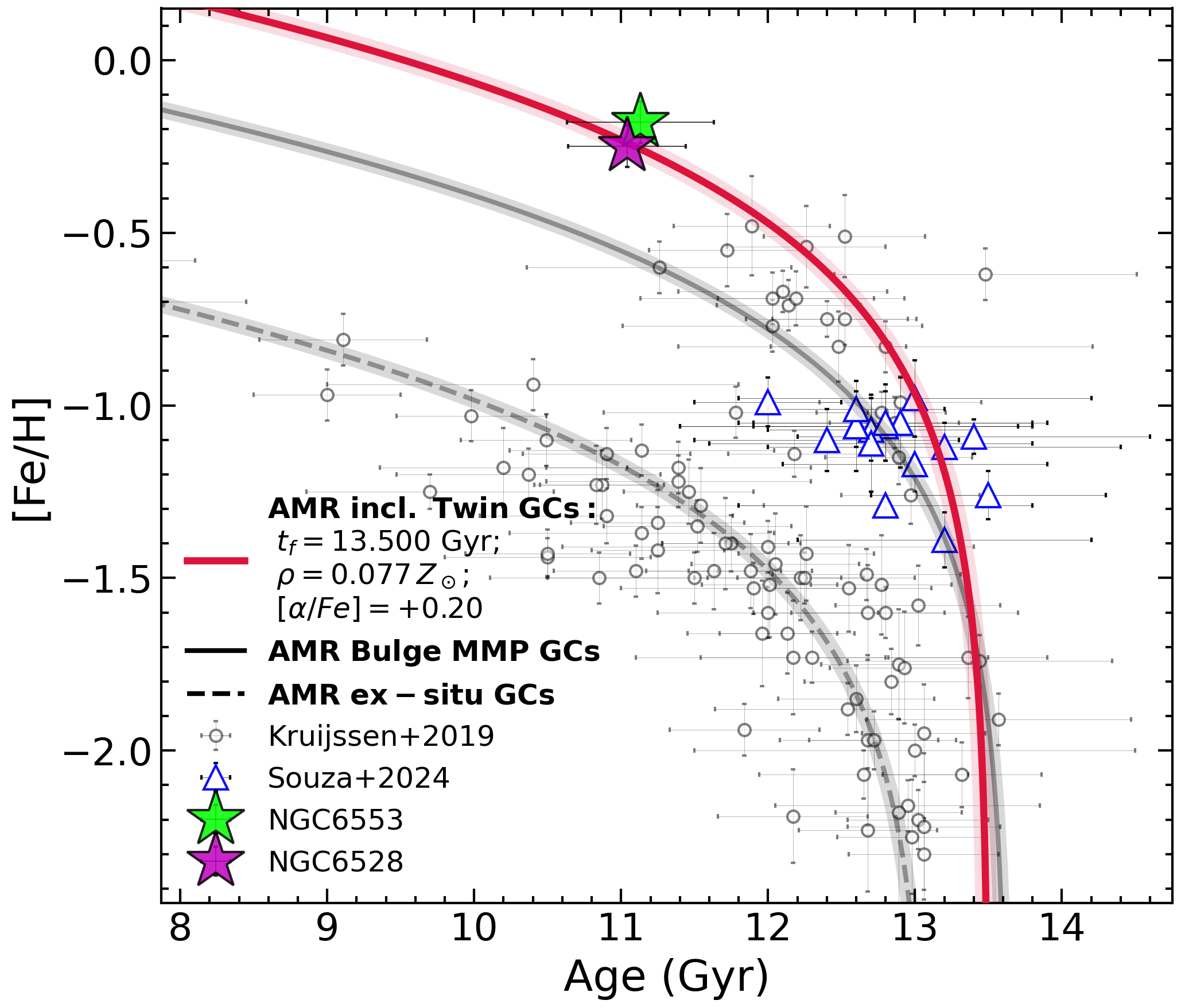}
    \caption{Age-metallicity relation (AMR) for the twin GCs. The red solid curve represents the AMR including the twin GCs, with a formation time of $t_f = 13.5$ Gyr, effective yields $\rho = 0.077 Z_\odot$, and $[\alpha/\text{Fe}] = +0.20$. The black solid and dashed curves represent the AMR for in-situ and ex-situ bulge metal-poor GCs, respectively. Open circles indicate the data from \cite{kruijssen19}, while blue triangles show the old moderately metal-poor clusters already presented in  \cite{souza24} for the metal-poor bulge GCs. The green and purple stars mark the positions of NGC 6553 and NGC 6528, respectively, highlighting their consistency with the AMR trend.}
    \label{fig:AMR}
\end{figure}

The theoretical AMR was obtained assuming a closed box and instantaneous recycling approximation (IRA). This approach is often used in the literature as a simple approximation of the in-situ and ex-situ branches \citep[e.g.][]{massari19,forbes20,souza24}. The two GCs add an important constraint to the bulge AMR. The AMR fitting resulted in a star formation starting $13.5$ Gyr and an effective yield of $0.077\,Z_\odot$. Compared to the values obtained in \cite{souza24} for the moderately metal-poor bulge GCs (MMPGCs), the values obtained in this work agree with the previous determination showing that the twin GCs improve the understanding of the formation and origin of the bulge GCs population since they fix the high end of the metallicity distribution function.

\begin{table*}
\caption{Mean abundance ratios for the sample clusters.}
%abundances of O,  odd-Z elements Na, Al,
%  alpha-elements Mg, Si, Ca, Ti, and heavy elements Zr, Ba, La, and Eu
%  from  \citet{alvesbrito06},  \citet{carretta01],  \citet{ernandes18}, and \citet{schiavon17}. }
\label{abund}
\scalefont{0.7}
\begin{flushleft}
\tabcolsep 0.12cm
%\begin{tabular}{l@{} {-} |  r@{}r@{} {-} | r@{} {-} | r@{}r@{}r@{}r@{} {-} | r@{}r@{}r@{}r@{}r@{}{-}| r@{}r@{}r@{}r@{}r@{}r@{} {-} |  }
%\begin{tabular}{l@{} {-} |  r@{}r@{} {-} | r@{} {-} | r@{}r@{}r@{} {-} | r@{}r@{}r@{}r@{}r@{} {-} | r@{}r@{}r@{}r@{} {-} |  }
\begin{tabular}{l r | r@{}r@{}r | r@{}r | r@{}r@{}r@{}r | r@{}r@{}r@{}r@{}r@{}r@{}r | r@{}r@{}r@{}r  | }

\noalign{\smallskip}
\hline\hline
\noalign{\smallskip}
author & [Fe/H] &  [C/Fe] & [N/Fe]  & [O/Fe] & [Na/Fe] & [Al/Fe] &  [Mg/Fe] & [Si/Fe] & [Ca/Fe]  & [Ti/Fe] &
 [Sc/Fe] & [V/Fe] & [Mn/Fe] & [Co/Fe] & [Ni/Fe] & [Cu/Fe] & [Zn/Fe] &
     [Zr/Fe] & [Ba/Fe] & [La/Fe] &  [Eu/Fe] \\ 
\noalign{\smallskip}
\hline
\noalign{\smallskip}
\multicolumn{19}{c}{\bf NGC~6528 }\\
\noalign{\smallskip}
\hline
\noalign{\smallskip}
Carretta    & +0.07   & ---   & ---   & +0.07            & +0.40 & ---     & +0.14 & +0.36 & +0.23  & +0.03 & $-$0.05 & $-$0.2  & $-$0.37 & ---     & ---   & --- & ---     & --- & +0.14 & --- & --- \\
Ernandes    & $-$0.10 & ---   & ---   & ---              & ---   & ---     & ---   & ---   & ---    &  ---  & $-$0.23 & $-$0.16 & $-$0.2  & $-$0.15 & ---   & --- & $-$0.26 & --- & ---   & --- & --- \\
Schiavon    & +0.00   & +0.01 & +0.16 & \phantom{-}+0.07 & +0.06 & $-$0.08 & +0.21 & +0.06 & $-$0.01&  ---  &  ---    & $-$0.16 & $-$0.04 & +0.03   & +0.05 & --- &  ---    & --- &  ---  & --- &  ---  \\
\noalign{\smallskip}
\hline
\noalign{\smallskip}
\multicolumn{19}{c}{\bf NGC~6553}  \\
\noalign{\smallskip}
\hline
\noalign{\smallskip}
Alves-Brito & \phantom{-}$-$0.20 & ---   & ---   & \phantom{-}+0.20 & +0.16 & +0.18 & +0.28 & +0.21 & +0.05 & $-$0.01 & +0.01& $-$0.1  & $-$0.26 & ---    & ---   & +0.14 & +0.08 & $-$0.67 & $-$0.28 & $-$0.11 & +0.10 \\
 Schiavon   & $-$0.17            & +0.08 & +0.18 & \phantom{-}+0.17 & +0.02 & +0.08 & +0.15 & +0.07 & +0.00 &  ---    &  --- & $-$0.12 &  +0.00  &  +0.08 & +0.06 & ---   &  ---  &  ---    &   ---   &  ---    &  --- \\    
\noalign{\smallskip}
\noalign{\hrule\smallskip}
%\noalign{\smallskip} 
%\hline 
\end{tabular}
\end{flushleft}
\end{table*}

\section{Concluding remarks}
\label{sec:disc}

We have carried out a photometric analysis of proper-motion-cleaned and differential reddening corrected
Colour-Magnitude Diagrams for NGC~6528 and NGC~6553. These much improved data supersede previous ones,
and in principle the analysis is more reliable than previous ones, 
mainly because of the wide time interval between the observations used for the proper motion cleaning.

We obtain ages of 11.04 and 11.1 Gyr for NGC~6528 and NGC~6553, and a solar metallicity
for both. The striking similarity of the two clusters in terms of age and metallicity is unique
among bulge globular clusters, in spite of their different locations and different total mass and
concentration
\citep[e.g.][]{trager95}.

For NGC~6553, the use of simultaneious fitting
with different wavelength baselines allowed to fix the total-to-selective absorption  R$_{\rm V}$ = 2.9,
therefore with a reliable distance from the Sun of 5.14 kpc.
For NGC~6528 we adopted a conservative total-to-selective absorption parameter R$_{\rm V}$ =
3.0, obtaining a distance from the Sun of 7.85 kpc.

These two metal-rich clusters are different from the moderatelly metal-poor bulge clusters,
that are older by about 2 Gyr. These combined studies allow to define an age-metallicity
relation for the Galactic bulge.

\begin{acknowledgements}
SO acknowledges the support of the University of Padova, DOR Ortolani 2020, Piotto 2021 and Piotto 2022, Italy and funded by the European Union – NextGenerationEU" RRF M4C2 1.1  n: 2022HY2NSX. "CHRONOS: adjusting the clock(s) to unveil the CHRONO-chemo-dynamical Structure of the Galaxy” (PI: S. Cassisi). 
 SOS acknowledges the DGAPA–PAPIIT grant IA103224 and the support from Dr. Nadine Neumayer's Lise Meitner grant from the Max Planck Society.
BB and EB acknowledge partial financial support from FAPESP, CNPq and CAPES - Financial code 001. 
 This research is based on observations made with the NASA/ESA \emph{Hubble} Space Telescope obtained from the Space Telescope Science Institute, which is operated by the Association of Universities for Research in Astronomy, Inc., under NASA contract NAS 5–26555.  The HST observations are associated with
 programmes GO-9453 (PI: Brown), GO-11664 (PI: Brown),  GO-10573 (PI: Mateo), GO-15232 (PI: Ferraro),
GO-16282 (PI: Correnti).

\end{acknowledgements}

\bibliographystyle{aa} % style aa.bst
\bibliography{n65286553}

\begin{appendix} %First appendix

The log of observations is given in Table~A\ref{obs}.

\begin{table}
\caption{Log of \emph{HST} observations.}
\label{obs}
\scalefont{0.7}
\begin{flushleft}
\tabcolsep 0.15cm
\begin{tabular}{lrrrrrrrrrrrrrrrrrrrrr}
%\noalign{\smallskip}
%\hline
\noalign{\smallskip}
\hline
\noalign{\smallskip}
PI & Proposal & Instrument & Filter & Exp.  &	DATE-OBS  \\  
\noalign{\smallskip}
\noalign{\hrule\smallskip}
%\noalign{\smallskip}
\multicolumn{6}{c}{\bf NGC~6528 }\\
\noalign{\smallskip}
\hline
\noalign{\smallskip}

Brown  & 9453     &  	ACS       & 	F606W    & 4.000000     & 	2002-08-06 \\	
Brown  & 9453     &  	ACS       & 	F606W    & 50.000000    & 	2002-08-06 \\	
Brown  & 9453     &  	ACS       & 	F606W    & 450.000000   & 	2002-08-06 \\	
Brown  & 9453     &  	ACS       & 	F814W    &  1.000000    & 	2002-08-06 \\	
Brown  & 9453     &  	ACS       & 	F814W    & 350.000000   & 	2002-08-06 \\	
Brown  & 9453     &  	ACS       & 	F814W    &  20.000000   & 	2002-08-06 \\	
Brown  & 11664    &  	WFC3      & 	F390W    & 40.000000    & 	2010-06-26 \\	
Brown  & 11664    &  	WFC3      & 	F390W    & 715.000000   & 	2010-06-26 \\	
Brown  & 11664    &  	WFC3      & 	F390W    & 348.000000   & 	2010-06-26 \\	
Brown  & 11664    &  	WFC3      & 	F390W    & 348.000000   & 	2010-06-26 \\	
Brown  & 11664    &  	WFC3      & 	F390W    & 715.000000   & 	2010-06-26 \\	
Brown  & 11664    &  	WFC3      & 	F390W    & 40.000000    & 	2010-06-26 \\	
Brown  & 11664    &  	WFC3      & 	F555W    & 1.000000     & 	2010-06-26 \\	
Brown  & 11664    &  	WFC3      & 	F555W    & 50.000000    & 	2010-06-26 \\	
Brown  & 11664    &  	WFC3      & 	F555W    & 665.000000   & 	2010-06-26 \\	
Brown  & 11664    &  	WFC3      & 	F555W    & 665.000000   & 	2010-06-26 \\	
Brown  & 11664    &  	WFC3      & 	F814W    & 370.000000   & 	2010-06-26 \\	
Brown  & 11664    &  	WFC3      & 	F814W    & 370.000000   & 	2010-06-26 \\	
Brown  & 11664    &  	WFC3      & 	F814W    & 1.000000     & 	2010-06-26 \\	
Brown  & 11664    &  	WFC3      & 	F814W    & 50.000000    & 	2010-06-26 \\	
\noalign{\smallskip}
\hline
\noalign{\smallskip}
\multicolumn{6}{c}{\bf NGC~6553}  \\
\noalign{\smallskip}
\hline
\noalign{\smallskip}
Mateo     &  10573    & ACS   & F435W   & 340.000000   & 2006-04-04 \\	
Mateo     &  10573    & ACS   & F435W   & 340.000000   & 2006-04-04 \\	
Mateo     &  10573    & ACS   & F435W   & 340.000000   & 2006-04-04 \\	
Mateo     &  10573    & ACS   & F555W   & 300.000000   & 2006-04-04 \\	
Mateo     &  10573    & ACS   & F814W   & 60.000000    & 2006-04-04 \\	
Ferraro   &  15232    & WFC3  & F555W   & 13.000000    & 2017-10-22 \\	
Ferraro   &  15232    & WFC3  & F555W   & 400.000000   & 2017-10-22 \\	
Ferraro   &  15232    & WFC3  & F555W   & 13.000000    & 2017-10-22 \\	
Ferraro   &  15232    & WFC3  & F555W   & 13.000000    & 2017-10-22 \\	
Ferraro   &  15232    & WFC3  & F555W   & 13.000000    & 2017-10-22 \\	
Ferraro   &  15232    & WFC3  & F555W   & 13.000000    & 2017-10-22 \\	
Ferraro   &  15232    & WFC3  & F555W   & 13.000000    & 2017-10-22 \\	
Ferraro   &  15232    & WFC3  & F814W   & 5.000000     & 2017-10-22 \\	
Ferraro   &  15232    & WFC3  & F814W   & 400.000000   & 2017-10-22 \\	
Ferraro   &  15232    & WFC3  & F814W   & 5.000000     & 2017-10-22 \\	
Ferraro   &  15232    & WFC3  & F814W   & 5.000000     & 2017-10-22 \\	
Ferraro   &  15232    & WFC3  & F814W   & 5.000000     & 2017-10-22 \\	
Ferraro   &  15232    & WFC3  & F814W   & 5.000000     & 2017-10-22 \\	
Ferraro   &  15232    & WFC3  & F814W   & 5.000000     & 2017-10-22 \\	
Correnti  &  16282    & WFC3  & F110W   & 299.231262   & 2020-10-19 \\	
Correnti  &  16282    & WFC3  & F110W   & 299.231262   & 2020-10-19 \\	
Correnti  &  16282    & WFC3  & F110W   & 299.231262   & 2020-10-19 \\	
Correnti  &  16282    & WFC3  & F110W   & 299.231262   & 2020-10-19 \\
Correnti  &  16282    & WFC3  & F160W   & 299.231262   & 2020-10-19 \\	
Correnti  &  16282    & WFC3  & F160W   & 299.231262   & 2020-10-19 \\	
Correnti  &  16282    & WFC3  & F160W   & 299.231262   & 2020-10-19 \\	
Correnti  &  16282    & WFC3  & F160W   & 299.231262   & 2020-10-19 \\	
\noalign{\smallskip} \hline \end{tabular}
\end{flushleft}
\end{table}

\end{appendix}

\end{document}